\documentclass[letterpaper,10pt,oneside,conference,final]{IEEEtran}

\usepackage[latin1]{inputenc}
\usepackage{graphicx}
\usepackage{color}
\usepackage{booktabs}
\usepackage{subfig}
\usepackage{tikz,pgfplots}
\usepackage{amssymb}
\usepackage{amsmath}
\usepackage{tabularx}
\usepackage{makecell}
\usetikzlibrary{spy}
\graphicspath{{figures/}}

\begin{document}
 	\IEEEoverridecommandlockouts
 	\title{Linear GFDM: A Low Out-of-band Emission Configuration for 5G Air Interface}
 	\author{Ivo Bizon Franco de Almeida and  Luciano Leonel Mendes \\
	 		{\small Instituto Nacional de Telecomunica\c{c}\~oes -- INATEL} \\
	 		{\small Santa Rita do Sapuca\'i, Brazil} \\
	 		{\small \texttt{ivobizon@gee.inatel.br,}} {\small \texttt{ luciano@inatel.br}}}
 	\maketitle

\begin{abstract}
	
	~This paper explores the flexibility of generalized frequency division multiplexing (GFDM) for achieving the same performance as filter bank multicarrier (FBMC). 
	We present a GFDM configuration where the good spectral containment of FBMC is also achieved through a modification in the GFDM transmission scheme. 
	For evaluating the scheme's performance, we estimate the bit error ratio (BER) under three channel models: \textit{i}) pure additive Gaussian AWGN; \textit{ii}) time-invariant frequency-selective; \textit{iii}) time-variant frequency-selective (doubly dispersive). The spectral containment is measured through the power spectral density. Furthermore, the complementary cumulative distribution function (CCDF) of the peak-to-average power ratio (PAPR) is also estimated. The Linear GFDM waveform shows identical performance when compared to FBMC in the above test scenarios.
	
\end{abstract}


\section{Introduction}

	\IEEEPARstart{R}{ecently} novel waveforms have been investigated to address the requirements of the fifth generation of cellular networks (5G). 
	Forecasts state that 5G will face challenging requirements in order to bring a vast new range of solutions and services \cite{waveformsfor5g}. 
	The envisioned connectivity scenarios and its basic physical layer (PHY) requirements are: \textit{i}) ultra reliable low latency (URLL) that requires very low end-to-end latency and reliability; \textit{ii}) enhanced mobile broadband (EMBB) that needs very high data throughput and low out-of-band (OOB) emissions; \textit{iii}) 5G for remote area access (RAA) that demands robustness against multipath propagation effects and low OOB emission in order to explore the vacant television white spaces (TVWS); \textit{iv}) Internet of Things (IoT)$/$Massive Machine Communications (MMC) that requires energy efficiency, support to a massive number of connections at a single radio base station and support to roughly synchronized transmissions.
	
	Regarding 5G's PHY standardization, there is an exciting discussion about which waveform is best suited for achieving the required performance among all mentioned scenarios. 
	The current 4G PHY employs orthogonal frequency division multiplexing (OFDM), which has been extensively studied and widely deployed in wired and wireless communication systems. 
	However, due to the wide range of requirements, some OFDM characteristics hinder its use in all 5G scenarios. 
	The necessity of maintaining strict synchronism between users and the radio base station, high OOB emission and inefficient use of the cyclic prefix (CP) are major OFDM shortcomings \cite{5gnow}. 
	Low OOB emission is an essential requirement to be fulfilled when one looks to the EMBB and RAA scenarios for example.
	
	The scope of requirements for 5G is much broader when compared to previous generations. 
	Therefore, an advanced multicarrier technique shall be employed in the next generation's PHY. 
	Currently, there are two main approaches described in literature. 
	On one hand, there are techniques which perform filtering subcarrier-wise, such as, filter bank multicarrier (FBMC) \cite{fbmcfor5g2} and generalized frequency division multiplexing (GFDM) \cite{gfdmfor5g}. 
	On the other hand, there are techniques that are closely related to OFDM and perform filtering subband-wise, e.g., universal filtered multicarrier (UFMC) \cite{ufmc} and filtered OFDM (F-OFDM) \cite{fofdm}. 
	These two latter techniques present an ease system migration and overall performance improvement when compared to OFDM. 
	Also, they do not lose spectral efficiency since the CP is not used. 
	However, the lack of CP leads to a waveform less robust to fading channels \cite{gfdmfor5g}.
	
	GFDM and FBMC are multicarrier techniques that abandon the strict orthogonality to achieve, for example, more spectral containment. 
	FBMC leads to ultra low OOB emission, which is very attractive in scenarios where the spectrum occupancy is highly dense. 
	The spectrum containment is achieved due to the subcarrier-wise linear filtering. 
	Furthermore, the spectrum containment also allows very low inference caused by roughly synchronized devices.
	Similarly to FBMC, GFDM also performs subcarrier-wise filtering, and its circular approach leads to a very flexible waveform. 
	It can be configured to satisfy low latency scenarios, and it shows the best spectral efficiency of all PHY candidates due to its efficient use of the CP \cite{waveformrace}.
		
	This paper shows that GFDM's flexibility can be exploited to achieve the same performance of FBMC regarding spectrum containment.
	Authors in \cite{datalinkstudy} mention that it is possible to configure GFDM to achieve the FBMC performance. 
	However, they do not show how to accomplish it.
	The main contribution of this paper is to precisely demonstrate how to modify the GFDM structure to generate the FBMC like signal.
	In addition, we present simulation results that confirm that the modifications do work.
	
	The rest of the paper is organized as follows: Section II presents the basic principles of non-orthogonal multicarrier techniques, specifically GFDM and FBMC. 
	Section III describes the GFDM configuration for achieving the FBMC performance. 
	Section IV presents the performance evaluation, comparing FBMC and OFDM to the modified GFDM under a common simulation setup. 
	Lastly, conclusions and final comments are presented in Section V.
	
\section{Background on GFDM and FBMC}

	GFDM and FBMC, differently from OFDM, are non-orthogonal waveforms, i.e., intrinsic intersymbol (ISI) and intercarrier (ICI) interference can arise. 
	However, by setting aside the orthogonality, a new degree of freedom is achieved according to the Balian-Low Theorem \cite{balianlow}. 
	As a result, a prototype filter with good time and frequency localization can be used for filtering subcarriers without decreasing spectral efficiency. 
	Hence, a spectrally efficient and contained waveform can be synthesized. 
	
	\subsection{GFDM principles}
	
	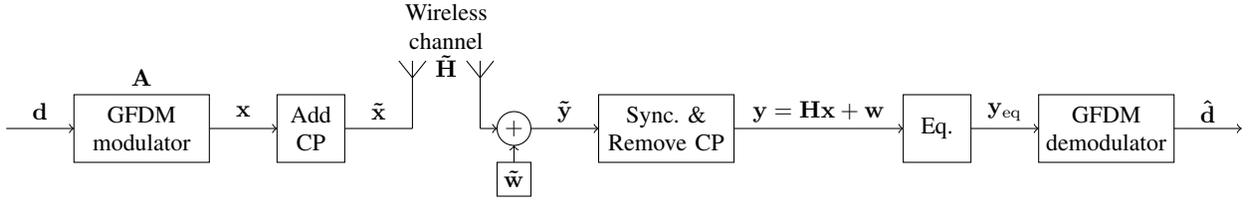
\begin{figure*}[t]
	\centering
	\begin{tikzpicture}[scale=0.9, every node/.style={scale=0.9}]
	\draw [->] (0,0)--(1,0);
	\node at (0.5,0.25) {$\mathbf{d}$};
	\draw (1,-0.5) rectangle (3,0.5);
	\node [align=center] at (2,0) {GFDM \\ modulator};
	\node at (2,0.75) {$\mathbf{A}$};
	\draw [->] (3,0)--(4,0);
	\node at (3.5,0.25) {$\mathbf{x}$};
	\draw (4,-0.5) rectangle (5,0.5);
	\node [align=center] at (4.5,0) {Add \\ CP};
	\draw (5,0)--(6,0);
	\node at (5.5,0.25) {$\mathbf{\tilde{x}}$};
	\draw (6,0)--(6,1);
	\draw (6,0.75)--(5.8,1);
	\draw (6,0.75)--(6.2,1);	
	\draw (7,0)--(7,1);
	\draw (7,0.75)--(7.2,1);
	\draw (7,0.75)--(6.8,1);		
	\node [align=center] at (6.5,1.3) {Wireless \\ channel \\ $\mathbf{\tilde{H}}$};
	\draw [->] (7,0)--(7.25,0);
	\draw (7.5,0) circle (0.25);
	\node at (7.5,0) {$+$};
	\draw [->] (7.5,-0.5)--(7.5,-0.25);
	\draw (7.25,-1) rectangle (7.75,-0.5);
	\node at (7.5,-0.75) {$\mathbf{\tilde{w}}$};
	\draw [->] (7.75,0)--(8.75,0);
	\node at (8.25,0.25) {$\mathbf{\tilde{y}}$};
	\draw (8.75,-0.5) rectangle (10.75,0.5);
	\node [align=center] at (9.75,0) {Sync. \&\\ Remove CP};
	\draw [->] (10.75,0)--(13.25,0);
	\node at (12,0.25) {$\mathbf{y = Hx + w}$};
	\draw (13.25,-0.5) rectangle (14.25,0.5);
	\node at (13.75,0) {Eq.};
	\draw [->] (14.25,0)--(15.25,0);
	\node at (14.75,0.25) {$\mathbf{y}_{\mathrm{eq}}$};
	\draw (15.25,-0.5) rectangle (17.25,0.5);
	\node [align=center] at (16.25,0) {GFDM \\ demodulator};
	\draw [->] (17.25,0)--(18.25,0);
	\node at (17.75,0.25) {$\mathbf{\hat{d}}$};
	\end{tikzpicture}	
	\caption{GFDM transceiver block diagram.}	
	\label{diagramgfdm}
	\end{figure*}
	
	GFDM is similar to FBMC in the sense that it is also based on filter bank theory, and it is built upon a prototype filter. 
	However, GFDM implements circular filtering, whereas FBMC implements linear filtering to shape subcarriers. 
	The GFDM waveform is composed by $M$ subsymbols circularly shifted in $M$ time-slots in $K$ different subcarriers. 
	Thus, a GFDM multicarrier frame carries $N=KM$ data symbols. 
	Since the waveform is obtained through circular filtering, it is possible to insert a CP to protect the frame against multipath propagation impairments. 
	Figure \ref{diagramgfdm} presents the GFDM block diagram \cite{book}. 
	
	The transmit vector $\mathbf{x}$, before the CP addition is given by
	\begin{equation} \label{eq1}
	\mathbf{x=Ad},
	\end{equation}
	where $\mathbf{d}$ represents the data symbol vector with $N$ complex-valued elements, which come from quadrature amplitude modulation (QAM) mapping. The $N$ versions of the prototype filter can be arranged for a transmit matrix $\mathbf{A}$, leading to
	\begin{equation} \label{eq2}
	\mathbf{A}=[\mathbf{g}_{0,0} \ \mathbf{g}_{1,0} \cdots \mathbf{g}_{K-1,0} \ \mathbf{g}_{0,1} \cdots \mathbf{g}_{K-1,M-1}]
	\end{equation}
	where $\mathbf{g}_{k,m}$ represents the prototype filter shifted to the $k$th subcarrier frequency and shifted to the $m$th subsymbol time slot. 
	
	The received vector after the CP removal is
	\begin{equation} \label{eq3}
	\mathbf{y=Hx+w},
	\end{equation}
	where $\mathbf{H}$ is the circulant channel matrix and it is obtained from the channel's impulse response. $\mathbf{w}$ is the additive white Gaussian noise (AWGN) vector with variance $\sigma^{2}$. For static-flat AWGN channel $\mathbf{H}$ is an identity matrix.
	
	At the receiver side, the transmitted data symbol vector can be recovered by a receiver matrix as follows
	\begin{equation} \label{eq4}
	\mathbf{\hat{d}=By},
	\end{equation}
	where $\mathbf{B}$ can be defined as a zero-forcing estimator by Eq. \ref{eq5} or a matched filter estimator by Eq. \ref{eq6}. It is important to point out that, the received vector $\mathbf{y}$ needs to be equalized first in these cases.
	Equation \ref{eq7} describes the receiver matrix for the minimum mean square error (MMSE) estimator. The MMSE estimator does not require previous equalization. However, it is a biased estimator, and requires a normalization of the estimated vector $\mathbf{\hat{d}}$ \cite{book}.
	\begin{equation} \label{eq5}
	\mathbf{B_{\mathrm{ZF}}=A}^{-1} 
	\end{equation}
	\begin{equation} \label{eq6}
	\mathbf{B_{\mathrm{MF}}=A}^{\mathrm{H}}
	\end{equation}
	\begin{equation} \label{eq7}
	\mathbf{B_{\mathrm{MMSE}}}=(\mathbf{R_{\mathrm{w}}}+\mathbf{A^{\mathrm{H}}}\mathbf{H^{\mathrm{H}}}\mathbf{HA})^{-1}\mathbf{A^{\mathrm{H}}}\mathbf{H^{\mathrm{H}}}
	\end{equation}
	$(\cdot)^{\mathrm{H}}$ represents the Hermitian operation and $\mathbf{R_{\mathrm{w}}}=\sigma^{2}\mathbf{I}_{N}$ represents the covariance matrix of the noise vector $\mathbf{w}$.	$\mathbf{I}_{N}$ represents the identity matrix of size $N\times N$, and $\mathbf{\tilde{x}}$ represents the transmit vector after the CP addition.
	
	\subsection{FBMC principles}
	
	FBMC was first proposed by Saltzberg \cite{saltzberg} and Chang \cite{chang} in around 1960, and it was brought back to be proposed as an alternative for the 5G PHY by Bellanger \cite{bellanger}.
	
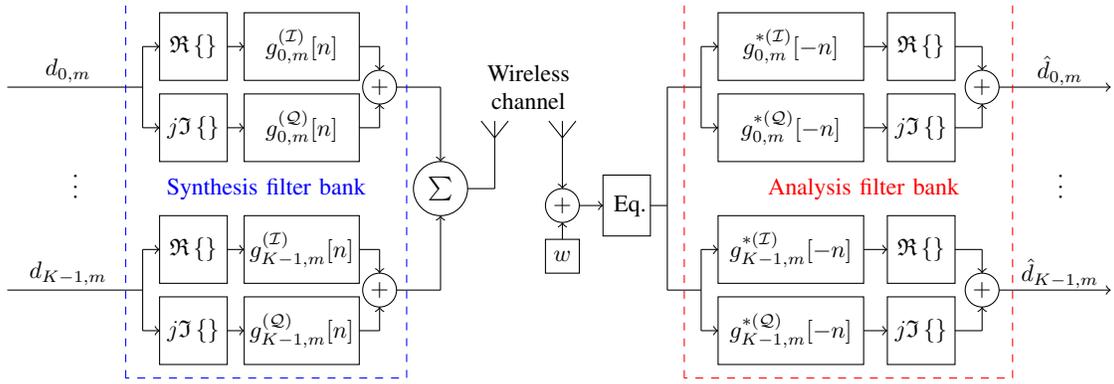
\begin{figure*}[t]
	\centering
	\begin{tikzpicture}[scale=0.9, every node/.style={scale=0.9}]
	\node at (1.9,0.25) {$d_{K-1,m}$};
	\draw (1,0)--(3,0);
	\draw (3,-0.6)--(3,0.6);
	\draw [->] (3,0.6)--(3.25,0.6);
	\draw [->] (3,-0.6)--(3.25,-0.6);	
	\draw (3.25,0.1) rectangle (4.25,1.1);
	\draw (3.25,-0.1) rectangle (4.25,-1.1);
	\node at (3.75,0.6) {$\mathfrak{R\left\lbrace \right\rbrace }$};
	\node at (3.75,-0.6) {$j\mathfrak{I\left\lbrace \right\rbrace }$};	
	\draw [->] (4.25,0.6)--(4.5,0.6);
	\draw [->] (4.25,-0.6)--(4.5,-0.6);	
	\draw (4.5,0.1) rectangle (6.2,1.1);
	\draw (4.5,-0.1) rectangle (6.2,-1.1);
	\node at (5.35,0.6) {$g_{K-1,m}^{\mathcal{(I)}}[n]$};
	\node at (5.35,-0.6) {$g_{K-1,m}^{\mathcal{(Q)}}[n]$};
	\draw (6.5,0) circle (0.25);
	\draw (6.2,0.6)--(6.5,0.6);
	\draw (6.2,-0.6)--(6.5,-0.6);
	\draw [->] (6.5,0.6)--(6.5,0.25);
	\draw [->] (6.5,-0.6)--(6.5,-0.25);	
	\node at (6.5,0) {$+$};
	\draw (6.75,0)--(7.4,0);
	\node [blue] at (4.825,1.5) {Synthesis filter bank};
	\node[text width = 3cm, align = center, rotate = -90] at (2,1.5) {$\cdots$};	
	\node at (1.9,3.25) {$d_{0,m}$};
	\draw (1,3)--(3,3);
	\draw (3,2.4)--(3,3.6);
	\draw [->] (3,3.6)--(3.25,3.6);
	\draw [->] (3,2.4)--(3.25,2.4);	
	\draw (3.25,3.1) rectangle (4.25,4.1);
	\draw (3.25,2.9) rectangle (4.25,1.9);
	\node at (3.75,3.6) {$\mathfrak{R\left\lbrace \right\rbrace }$};
	\node at (3.75,2.4) {$j\mathfrak{I\left\lbrace \right\rbrace }$};	
	\draw [->] (4.25,3.6)--(4.5,3.6);
	\draw [->] (4.25,2.4)--(4.5,2.4);	
	\draw (4.5,3.1) rectangle (6.2,4.1);
	\draw (4.5,2.9) rectangle (6.2,1.9);
	\node at (5.35,3.6) {$g_{0,m}^{\mathcal{(I)}}[n]$};
	\node at (5.35,2.4) {$g_{0,m}^{\mathcal{(Q)}}[n]$};
	\draw (6.5,3) circle (0.25);
	\draw (6.2,3.6)--(6.5,3.6);
	\draw (6.2,2.4)--(6.5,2.4);
	\draw [->] (6.5,3.6)--(6.5,3.25);
	\draw [->] (6.5,2.4)--(6.5,2.75);	
	\node at (6.5,3) {$+$};
	\draw (6.75,3)--(7.4,3);		
	\draw [blue,dashed] (2.75,-1.3) rectangle (6.9,4.3);
	\draw (7.4,1.5) circle (0.4);
	\node at (7.4,1.5) {$\sum$};
	\draw [->] (7.4,3)--(7.4,1.9);
	\draw [->] (7.4,0)--(7.4,1.1);	
	\draw (7.8,1.5)--(8.2,1.5);
	\draw (8.2,1.5)--(8.2,2.5);
	\draw (8.2,2.25)--(8,2.5);
	\draw (8.2,2.25)--(8.4,2.5);
	\draw [<-] (9.2,1.5)--(9.2,2.5);
	\draw (9.2,2.25)--(9,2.5);
	\draw (9.2,2.25)--(9.4,2.5);
	\node [align=center] at (8.7,3) {Wireless \\ channel};
	\draw (9.2,1.25) circle (0.25);	
	\node at (9.2,1.25) {$+$};
	\draw [->] (9.2,0.75)--(9.2,1);
	\draw (8.95,0.25) rectangle (9.45,0.75);
	\node at (9.2,0.5) {$w$};
	\draw [->] (9.45,1.25)--(9.8,1.25);
	\draw (9.8,0.8) rectangle (10.5,1.7);
	\node at (10.2,1.25) {Eq.};
	\draw (10.5,1.25)--(10.75,1.25);
	\draw (10.75,0)--(10.75,3);
	\draw (10.75,0)--(11.25,0);
	\draw (11.25,-0.6)--(11.25,0.6);
	\draw [->] (11.25,0.6)--(11.5,0.6);
	\draw [->] (11.25,-0.6)--(11.5,-0.6);
	\draw (11.5,0.1) rectangle (13.65,1.1);
	\draw (11.5,-1.1) rectangle (13.65,-0.1);
	\node at (12.575,0.6) {$g_{K-1,m}^{*\mathcal{(I)}}[-n]$};
	\node at (12.575,-0.6) {$g_{K-1,m}^{*\mathcal{(Q)}}[-n]$};
	\draw [->] (13.65,0.6)--(14,0.6);
	\draw [->] (13.65,-0.6)--(14,-0.6);	
	\draw [->] (13.65,3.6)--(14,3.6);
	\draw [->] (13.65,2.4)--(14,2.4);	
	\draw (14,3.1) rectangle (15,4.1);
	\draw (14,2.9) rectangle (15,1.9);
	\node at (14.5,3.6) {$\mathfrak{R\left\lbrace \right\rbrace }$};
	\node at (14.5,2.4) {$j\mathfrak{I\left\lbrace \right\rbrace }$};	
	\draw (14,-1.1) rectangle (15,-0.1);
	\draw (14,0.1) rectangle (15,1.1);
	\node at (14.5,0.6) {$\mathfrak{R\left\lbrace \right\rbrace }$};
	\node at (14.5,-0.6) {$j\mathfrak{I\left\lbrace \right\rbrace }$};		
	\draw (15,0.6)--(15.4,0.6);
	\draw (15,-0.6)--(15.4,-0.6);	
	\draw (15,3.6)--(15.4,3.6);	
	\draw (15,2.4)--(15.4,2.4);		
	\draw (15.4,0) circle (0.25);
	\node at (15.4,0) {$+$};
	\draw (15.4,3) circle (0.25);	
	\node at (15.4,3) {$+$};
	\draw [->] (15.4,2.4)--(15.4,2.75);
	\draw [->] (15.4,3.6)--(15.4,3.25);	
	\draw [->] (15.4,-0.6)--(15.4,-0.25);
	\draw [->] (15.4,0.6)--(15.4,0.25);	
	\draw [->] (15.65,0)--(17.3,0);
	\draw [->] (15.65,3)--(17.3,3);	
	\node at (16.55,3.25) {$\hat{d}_{0,m}$};	
	\node at (16.55,0.25) {$\hat{d}_{K-1,m}$};
	\draw (10.75,3)--(11.25,3);
	\draw (11.25,2.4)--(11.25,3.6);
	\draw [->] (11.25,3.6)--(11.5,3.6);
	\draw [->] (11.25,2.4)--(11.5,2.4);
	\draw (11.5,3.1) rectangle (13.65,4.1);
	\draw (11.5,1.9) rectangle (13.65,2.9);
	\node at (12.575,3.6) {$g_{0,m}^{*\mathcal{(I)}}[-n]$};
	\node at (12.575,2.4) {$g_{0,m}^{*\mathcal{(Q)}}[-n]$};
	\node[text width = 3cm, align = center, rotate = -90] at (16.55,1.5) {$\cdots$};	
	\draw [dashed, red] (11,-1.3) rectangle (15.85,4.3);
	\node [red] at (13.65,1.5) {Analysis filter bank};
	\end{tikzpicture}
	\caption{Discrete time FBMC-OQAM transceiver block diagram.}
	\label{diagramfbmc}
	\end{figure*}

	For the FBMC notation, let $K$ represent the total number of subcarriers, $Lp$ the prototype filter length in samples, $\Theta$ the overlapping factor and $T$ the QAM data symbol time spacing. 
	The prototype filter length is set to $Lp=\Theta K+1$. 
	The overlapping factor corresponds to the number of data symbol periods in the prototype filter length. 
	For example, if the overlapping factor is set to $\Theta=4$, then the prototype filter length is $4$ times longer than the data symbol period. 
	Therefore, it means that the pulse shaped data symbols overlap in time domain.
	Half Nyquist prototype filters with good frequency localization are preferred since they do not cause ISI at the sampling instant and minimize the ICI among subcarriers.
	A possible implementation of the FBMC-OQAM transceiver structure is depicted in Fig. \ref{diagramfbmc} \cite{fbmcfor5g}.
		
	Each subcarrier is filtered with a version of the prototype filter shifted in the correspondent subcarrier frequency.  
	Granted that subcarriers overlap in frequency domain only with its neighbors, maximum spectral efficiency can be achieved through Offeset-QAM (OQAM) mapping. 
	The prototype filter proposed in \cite{aprimer}, extensively used in FBMC context, shows the desired characteristics while keeping only real coefficients, which is essential for employing OQAM mapping.
	 
	OQAM mapping can vanish neighboring ICI by shifting the real part of a QAM symbol by $T/2$ in relation to the imaginary part. 
	Therefore, orthogonality is restored and maximum spectral efficiency is achieved at the expense of an increase in the sampling rate by a factor of $2$. 
	
	The discrete-time FBMC-OQAM transmit signal is obtained by
	\begin{multline} \label{fbmcwave}
		x[n] = \sum_{m=-\infty}^{+\infty} \sum_{k=0}^{K-1} \mathfrak{R} \left\lbrace  d_{k,m} \right\rbrace  g^{\mathcal{(I)}}_{k,m}\left[n\right] \\ + j\sum_{m=-\infty}^{+\infty} \sum_{k=0}^{K-1} \mathfrak{I}\left\lbrace d_{k,m} \right\rbrace  g^{\mathcal{(Q)}}_{k,m}[n],
	\end{multline}
	where $d_{k,m}$ is the complex QAM data symbol that is transmitted over the $k$th subcarrier in the $m$th time-slot.
	
	$g^{\mathcal{(I)}}_{k,m}[n]$ and $g^{\mathcal{(Q)}}_{k,m}[n]$ compose the synthesis filter bank, and they are respectively given by
	\begin{equation} \label{gi}
		g^{\mathcal{(I)}}_{k,m}[n] = p\left[ n-mK\right] e^{j2\pi \frac{k}{K}n} e^{j\frac{\pi}{2}k}
	\end{equation}
	\begin{equation} \label{gq}
	g^{\mathcal{(Q)}}_{k,m}[n] = p\left[ n-\left( m+\frac{1}{2}\right) K\right] e^{j2\pi \frac{k}{K}n} e^{j\frac{\pi}{2}k} ,
	\end{equation}
	where $p[n]$ is the prototype filter impulse response.
	The received signal passes through the analysis filter bank, and the estimated data symbols $\hat{d}_{k,m}$ are obtained.
	
\section{GFDM as FBMC}

	\begin{figure*}[t!]
		\centering
		\begin{tikzpicture}[scale=0.9, every node/.style={scale=0.9}]
		\draw (0,0)--(1,0);
		\node at (0.5,0.25) {$\mathbf{d}$};
		\draw (1,-0.6)--(1,0.6);
		\draw [->] (1,0.6)--(1.25,0.6);
		\draw [->] (1,-0.6)--(1.25,-0.6);		
		\draw (1.25,-1.1) rectangle (2.25,-0.1);
		\node at (1.75,-0.6) {$j\mathfrak{I}\left\lbrace  \right\rbrace $};
		\draw (1.25,1.1) rectangle (2.25,0.1);
		\node at (1.75,0.5) {$\mathfrak{R}\left\lbrace  \right\rbrace $};	
		\draw [->] (2.25,0.6)--(2.5,0.6);
		\draw [->] (2.25,-0.6)--(2.5,-0.6);
		\draw (2.5,-1.1) rectangle (3.5,-0.1);
		\node at (3,-0.6) {$\mathbf{A}_{\mathrm{q}}^{\mathrm{(L)}}$};
		\draw (2.5,1.1) rectangle (3.5,0.1);
		\node at (3,0.6) {$\mathbf{A}_{\mathrm{i}}^{\mathrm{(L)}}$};		
		\draw (3.5,0.6)--(4,0.6);
		\draw (3.5,-0.6)--(4,-0.6);
		\draw [->] (4,-0.6)--(4,-0.35);
		\draw [->] (4,0.6)--(4,0.35);		
		\draw (4,0) circle (0.35);
		\node at (4,0) {$\sum$};
		\draw (4.35,0)--(5,0);
		\node at (4.625,0.2) {$\mathbf{x}$};
		\draw (5,0)--(5,1);
		\draw (5,0.75)--(5.2,1);
		\draw (5,0.75)--(4.8,1);
		\node [align=center] at (5.5,1.5) {Wireless \\ channel};
		\draw (6,0)--(6,1);
		\draw (6,0.75)--(5.8,1);
		\draw (6,0.75)--(6.2,1);
		\draw [->] (6,0)--(6.25,0);
		\draw (6.5,0) circle (0.25);
		\node at (6.5,0) {$+$};
		\draw [->] (6.5,-0.5)--(6.5,-0.25);
		\draw (6.25,-1) rectangle (6.75,-0.5);
		\node at (6.5,-0.75) {$\mathbf{w}$};
		\draw [->] (6.75,0)--(7.5,0);
		\node at (7.125,0.25) {$\mathbf{y}$};
		\draw (7.5,-0.5) rectangle (8.5,0.5);
		\node at (8,0) {Eq.};
		\draw (8.5,0)--(9.5,0);
		\node at (9,0.25) {$\mathbf{y}_{\mathrm{eq}}$};
		\draw (9.5,-0.6)--(9.5,0.6);
		\draw [->] (9.5,0.6)--(9.75,0.6);
		\draw [->] (9.5,-0.6)--(9.75,-0.6);
		\draw (9.75,0.1) rectangle (10.75,1.1);
		\draw (9.75,-0.1) rectangle (10.75,-1.1);
		\node at (10.25,-0.6) {$\mathbf{B}_{\mathrm{q}}^{\mathrm{(L)}}$};
		\node at (10.25,0.6) {$\mathbf{B}_{\mathrm{i}}^{\mathrm{(L)}}$};
		\draw [->] (10.75,0.6)--(11,0.6);
		\draw [->] (10.75,-0.6)--(11,-0.6);
		\draw (11,-0.1) rectangle (12,-1.1);
		\draw (11,0.1) rectangle (12,1.1);
		\node at (11.5,-0.6) {$j\mathfrak{I}\left\lbrace  \right\rbrace $};
		\node at (11.5,0.6) {$\mathfrak{R}\left\lbrace  \right\rbrace $};	
		\draw (12,0.6)--(12.5,0.6);
		\draw (12,-0.6)--(12.5,-0.6);
		\draw [->] (12.5,0.6)--(12.5,0.35);
		\draw [->] (12.5,-0.6)--(12.5,-0.35);
		\draw (12.5,0) circle (0.35);
		\node at (12.5,0) {$\sum$};
		\draw [->] (12.85,0)--(13.85,0);
		\node at (13.35,0.25) {$\mathbf{\hat{d}}$};
		\end{tikzpicture}
		\caption{Linear GFDM transceiver block diagram.}
		\label{diagrama_linear_gfdm}
	\end{figure*}
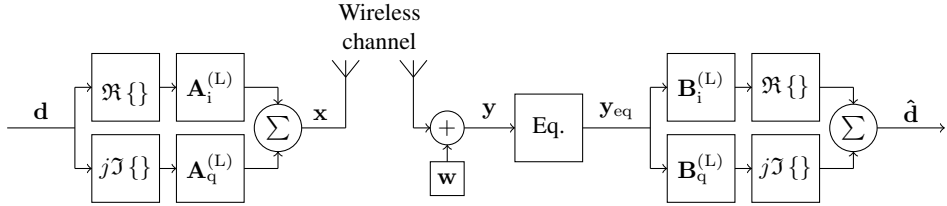

\begin{figure}[h]
			\subfloat[Absolute value of $\mathbf{A}_{\mathrm{i}} $.]{\includegraphics[width=7.6cm,height=5.5cm]{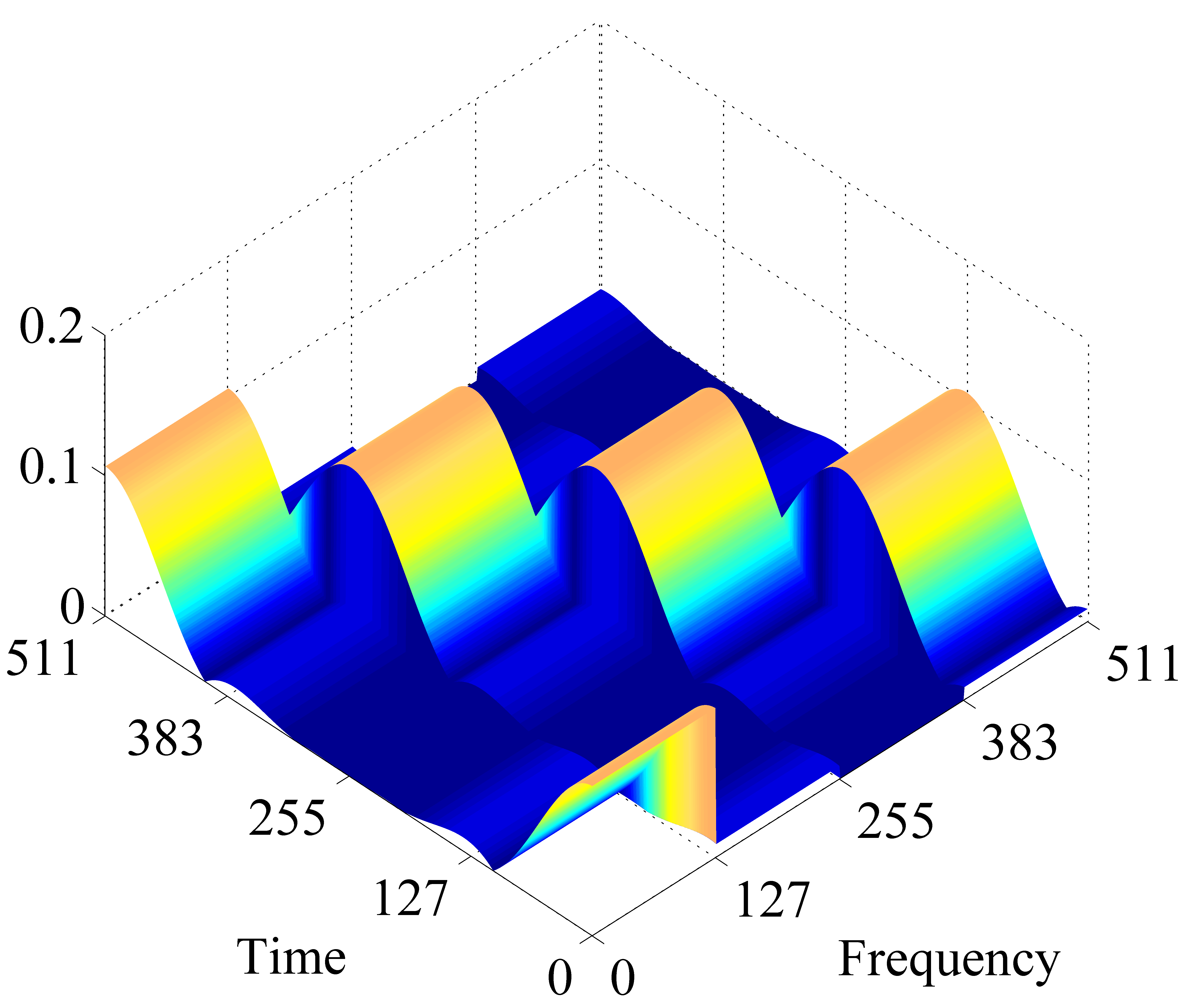}} \quad
			\subfloat[Absolute value of $\mathbf{A}_{\mathrm{q}} $.]{\includegraphics[width=7.6cm,height=5.5cm]{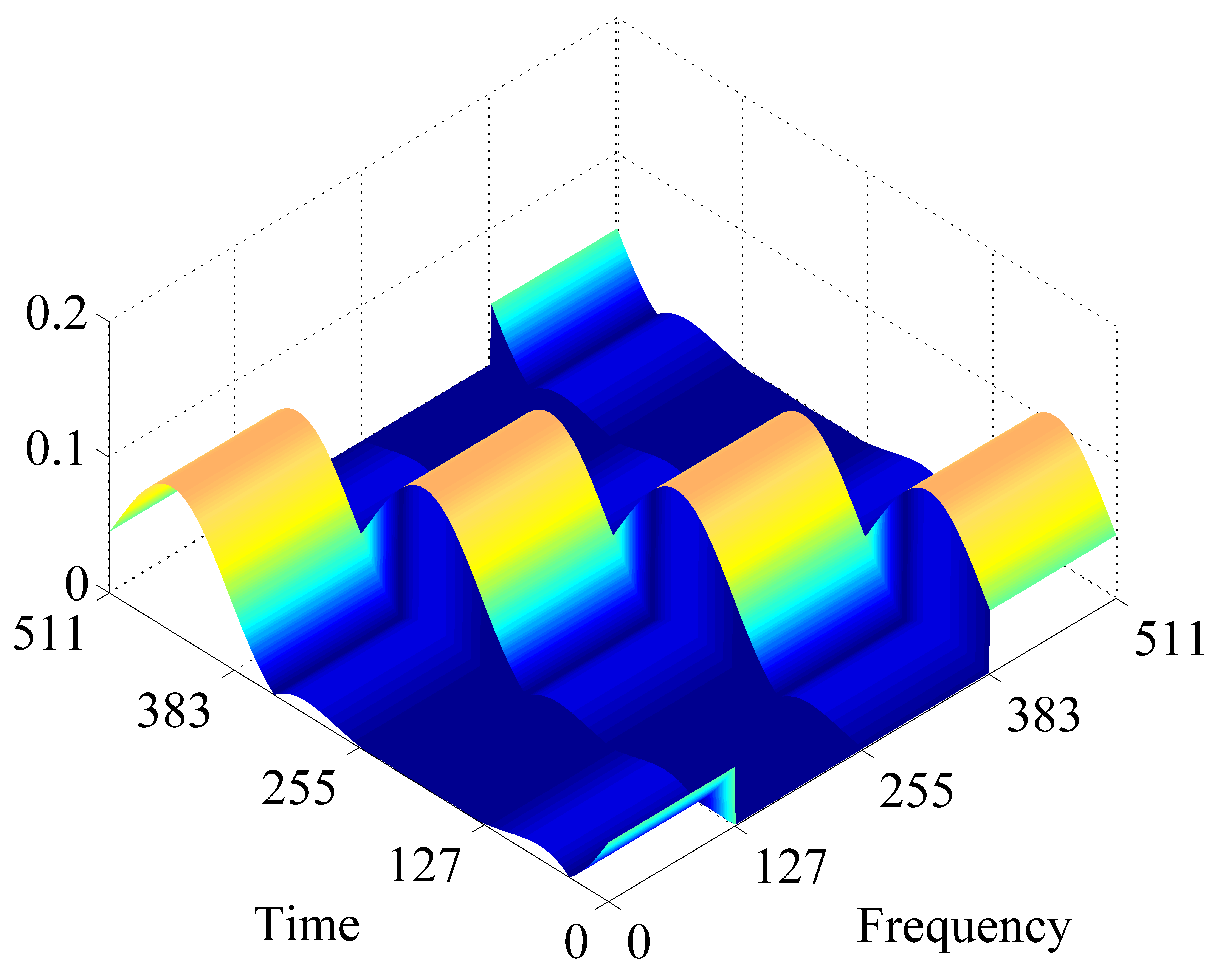}}
			\centering
			\caption{Absolute value of the transmit matrices for $M = 4$ subsymbols, $K = 128$ subcarriers, PHYDYAS prototype filter \cite{aprimer}. In the time-domain OQAM configuration $\mathbf{A}_{\mathrm{q}}$ is half subsymbol shifted in relation to $\mathbf{A}_{\mathrm{i}}$.}
			\label{modmat}
	\end{figure}

	In GFDM context, OQAM can be employed by creating two modulation matrices, $\mathbf{A}_{\mathrm{i}}$ and $\mathbf{A}_{\mathrm{q}}$, where one is half subsymbol circularly shifted in relation to the other \cite{oqamgfdm}. $\mathbf{A}_{\mathrm{i}}$ modulates the real part of a QAM symbol, whereas $\mathbf{A}_{\mathrm{q}}$ modulates the imaginary part. These matrices are obtained similarly to Eq. \ref{eq2}. Figure \ref{modmat} illustrates absolute value of the GFDM transmit matrices for the OQAM configuration.
		
	The transmit vector is obtained through a matrix multiplication among the real and imaginary parts of the data symbol vector and the transmit matrices as follows
	\begin{equation}
		\mathbf{x} = \mathbf{A}_{\mathrm{i}} \mathfrak{R} \left\lbrace  \mathbf{d} \right\rbrace + j \mathbf{A}_{\mathrm{q}} \mathfrak{I} \left\lbrace  \mathbf{d} \right\rbrace
	\end{equation}
	
	After equalization at the receiver side, the estimated data vector is given by
	\begin{equation}\label{rxs_gfdm_oqam}
		\mathbf{\hat{d}} = \mathfrak{R}\left\lbrace \mathbf{B}_{\mathrm{i}}\mathbf{y_{\mathrm{eq}} } \right\rbrace + j \mathfrak{I}\left\lbrace \mathbf{B}_{\mathrm{q}}\mathbf{y_{\mathrm{eq}}} \right\rbrace,
	\end{equation}
	where $\mathbf{B}_{\mathrm{i}}$ and $\mathbf{B}_{\mathrm{q}}$ represent the matched filter demodulation matrix, and are respectively given by
	\begin{gather}\label{linear_demod_mats}
		\mathbf{B}_{\mathrm{i}} = \mathbf{A}_{\mathrm{i}}^{\mathrm{H}} \\
		\mathbf{B}_{\mathrm{q}} = \mathbf{A}_{\mathrm{q}}^{\mathrm{H}}
	\end{gather}
	
	However, as one may see in Fig. \ref{modmat} the circular shifting of the prototype filter leads to abrupt transitions at the edges of the transmit matrices. 
	Hence, OOB emission is rather elevated due to discontinuities in the transmitted waveform. 
	
	Flexibility in the GFDM transmit matrix is exploited in order to generate the Linear GFDM waveform, which performs identically to FBMC. 
	We propose zero padding the prototype filter in order to get a linear filtering behavior from the transmit matrices. 
	Figure \ref{diagrama_linear_gfdm} presents a block diagram that describes the transmission and reception of Linear GFDM, where $\mathbf{A}^{\mathrm{(L)}}_{\mathrm{i}} $ denotes the modified transmit matrix that modulates the real part the data vector, and $\mathbf{A}^{\mathrm{(L)}}_{\mathrm{q}}$ the modified transmit matrix that modulates the imaginary part.	
	The matrices $\mathbf{B}_{\mathrm{i}}^{\mathrm{(L)}}$ and $\mathbf{B}_{\mathrm{q}}^{\mathrm{(L)}}$ respectively represent the matched filter receiver matrices and are obtained similarly to Eq. \ref{linear_demod_mats}. .
	
\begin{figure}[h]
		\subfloat[Absolute value of $\mathbf{A}^{\mathrm{(L)}}_{\mathrm{i}} $.]{\includegraphics[width=7.6cm,height=5.5cm]{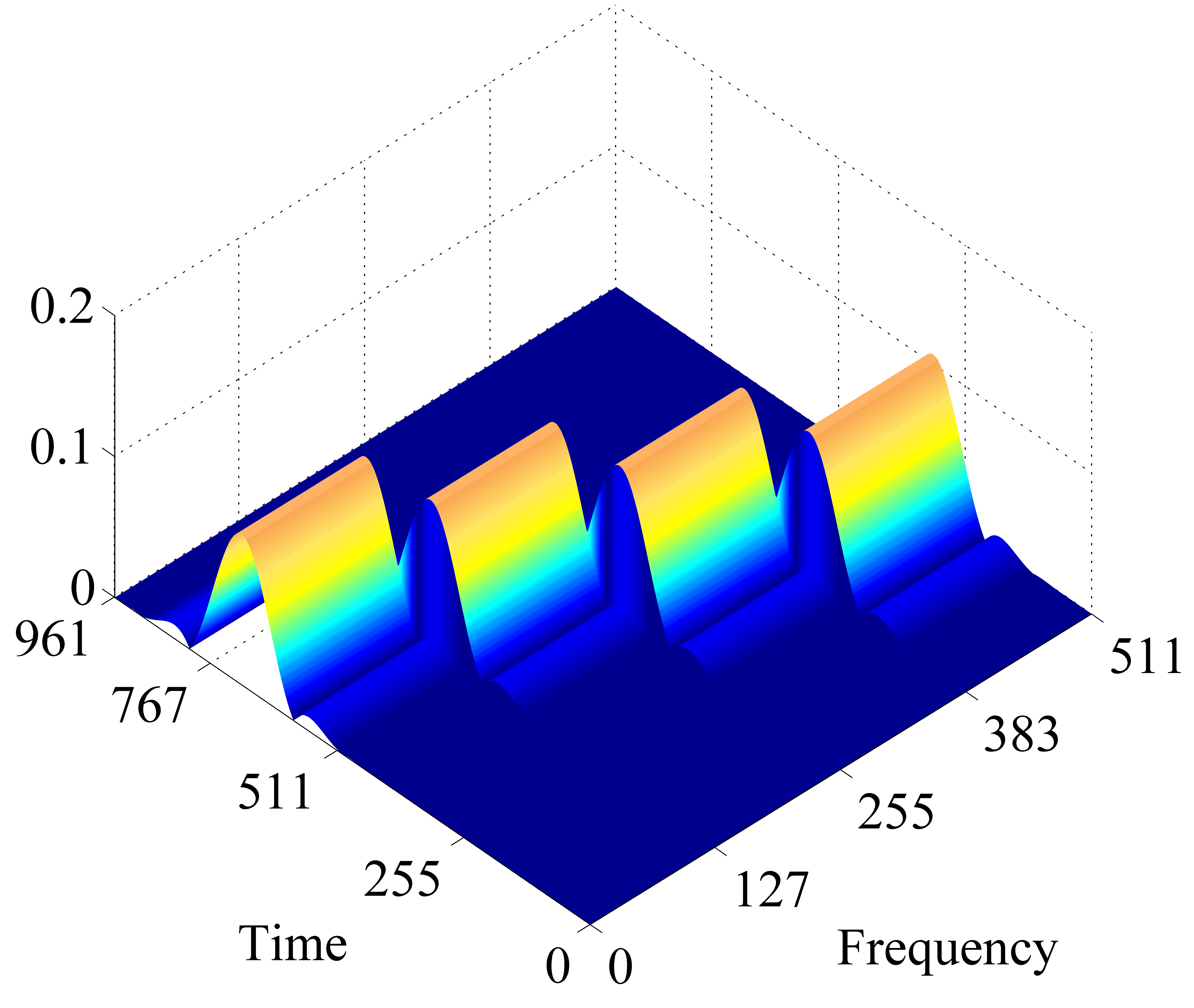}} \quad
		\subfloat[Absolute value of $\mathbf{A}^{\mathrm{(L)}}_{\mathrm{q}} $.]{\includegraphics[width=7.6cm,height=5.5cm]{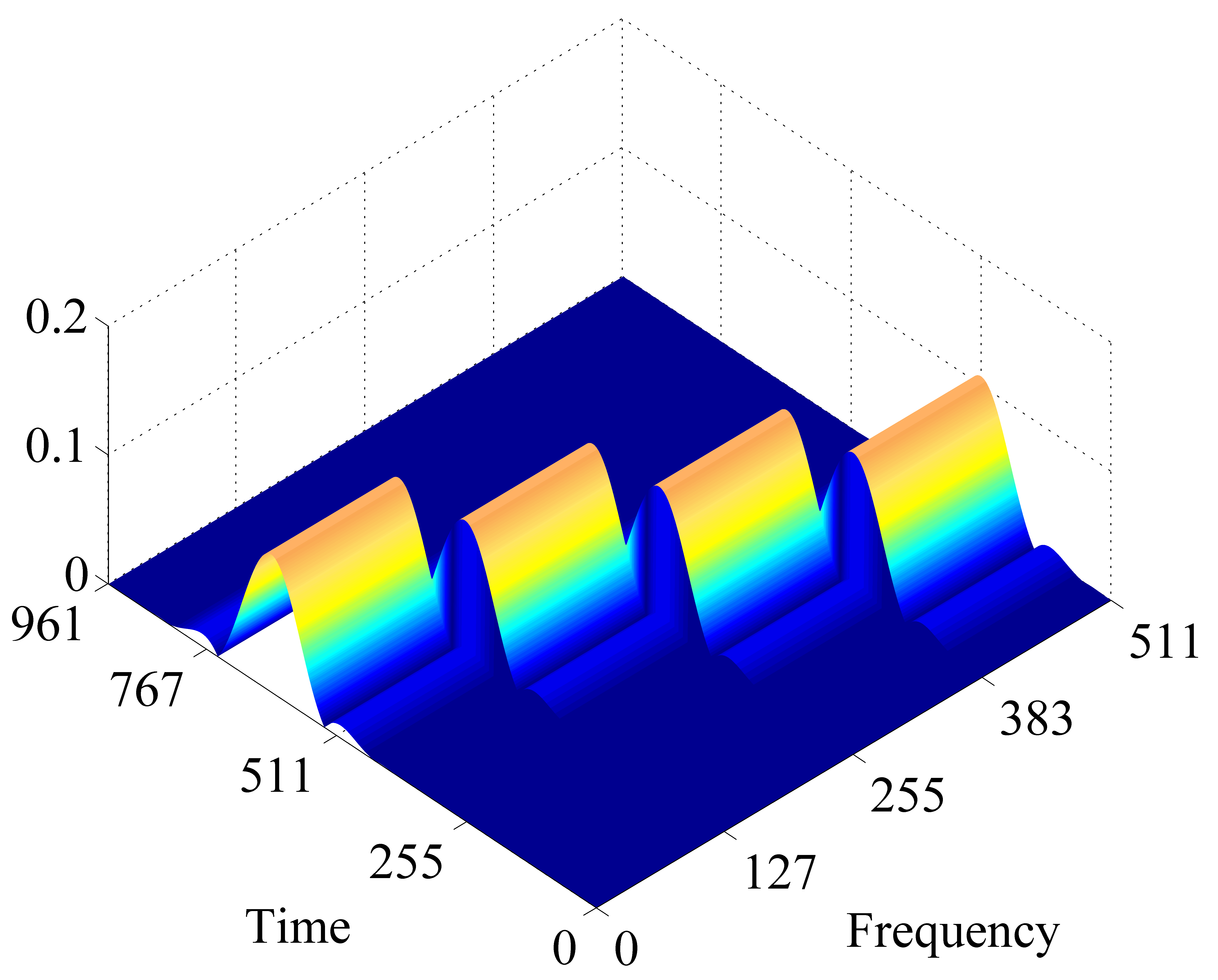}}
		\centering
		\caption{Absolute value of the modified transmit matrices. $M = 4$ subsymbols, $K = 128$ subcarriers, PHYDYAS prototype filter.}
		\label{modmatlinear}
	\end{figure}

	Let $L_Z$ denote the zero-valued sequence length that is padded to the prototype filter.
	If the prototype filter is zero padded according with Eq. \ref{eq33}, then no abrupt transitions are observed at the transmit matrix and linear filtering behavior is observed. 
	\begin{equation} \label{eq33}
		L_Z = K(M-1)+1,
	\end{equation}
	where $M$ is the GFDM notation for the number of subsymbols and $K$ is the number of subcarriers.
	The unity is added to ensure no inflections on the transmitted waveform. 
	However, as the matrix $\mathbf{A}_{\mathrm{q}} $ is $K/2$ samples shifted in relation to $\mathbf{A}_{\mathrm{i}} $ it is necessary to add $K/2$ to Eq. \ref{eq33}, leading to
	\begin{equation}
		L_Z = KM-K/2+1,
	\end{equation}
	by doing so, there are no abrupt transitions in either matrices.	
	Figure \ref{modmatlinear} illustrates the absolute value of the modified transmit matrix for achieving the linear filtering behavior.
	
\section{Performance evaluation}

	For evaluating the performance, numerical simulations using MATLAB were carried out.
	Firstly, the bit error ratio (BER) from Linear GFDM, FBMC-OQAM and OFDM are presented, and the theoretical OFDM bit error probability is also shown \cite{whtofdmleonel}\cite{alouini}.
	The BER performance is evaluated under three channel models: \textit{i}) pure additive Gaussian AWGN; \textit{ii}) time-invariant frequency-selective with eight taps; \textit{iii}) time-variant frequency-selective channel with four taps.
	Secondly, the power spectral density (PSD) from Linear GFDM, FBMC-OQAM, circular GFDM-OQAM and OFDM are compared. 
	Finally, the cumulative distribution function (CCDF) of the peak-to-average power ratio (PAPR) is presented.
	Table \ref{tabela1} shows the waveforms' simulation parameters and Table \ref{tabela2} shows the channel models used for the simulations.
	
	\begin{table}[h]
		\centering
		\caption{Waveform related simulation parameters}
		\label{tabela1}
		\renewcommand{\arraystretch}{1.5}
		\begin{tabularx}{\columnwidth}{llll}
			\toprule[0.9pt] 
			Parameter                & Linear GFDM            & FBMC-OQAM              & OFDM     \\ 
			\midrule 
			Mapping                  & 16-QAM                 & 16-QAM                 & 16-QAM   \\ 
			Prototype filter         & PHYDYAS \cite{aprimer} & PHYDYAS 			   & Rect     \\ 
			Number of subcarriers    & 128                    & 128                    & 512      \\ 
			Number of subsymbols     & 4                      & -                      & 1        \\ 
			Overlapping factor       & -                      & 4                      & -        \\ 
			Data symbols per frame   & 512                    & 512                    & 512      \\
			\bottomrule[0.9pt]	
		\end{tabularx}
	\end{table}

	\begin{table}[h]
		\centering
		\caption{Channel models}
		\label{tabela2}
		\renewcommand{\arraystretch}{1.5}
		\renewcommand\cellgape{\Gape[3pt]}
		\begin{tabularx}{\columnwidth}{ll}
			\toprule[0.9pt] 
			Channel                                              & Impulse response  					 \\
			\midrule 
			Pure AWGN                                     & $\mathbf{h}_{\mathrm{AWGN}} = 1$  					 \\ 
			\makecell[lc]{Time-invariant \\ frequency-selective}    & $\mathbf{h}_{\mathrm{TIFS}} = [1\;0\; 0\; 0\; 0.4\; 0\; 0\; 0.2]$   \\ 
			\makecell[lc]{Time-variant \\ frequency-selective}     & \makecell[lc]{ $ \mathbf{h}_{\mathrm{TVFS}} =  \left[ \frac{1}{\sqrt{2}} \mathbf{r}_{0} \; \frac{0}{\sqrt{2}} \mathbf{r}_{1} \; \frac{0.01^{2} }{\sqrt{2}} \mathbf{r}_{2} \; \frac{0.02^{2}}{\sqrt{2}} \mathbf{r}_{3} \right]  $ \\ \Gape[5pt]{ $ \mathbf{r}_{n}\sim\mathcal{CN}(0,1) $ } }  \\
			\bottomrule[0.9pt]	
		\end{tabularx}
	\end{table}
\subsection{BER performance}
	
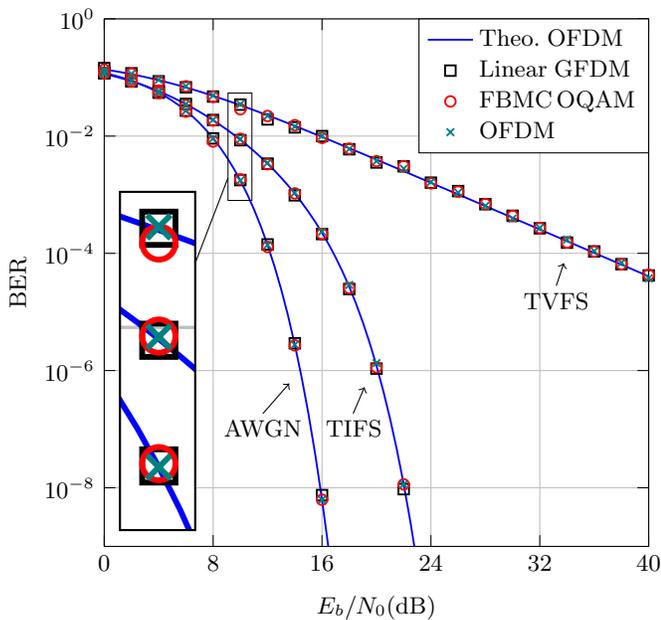
\begin{figure}[h]
	\centering
	\begin{tikzpicture}[scale=1, every node/.style={scale=1}, spy using outlines={rectangle, magnification=3.15, width=1cm, height=4.5cm}, connect spies, font=\small]
	\begin{axis}[%
	width=8.82cm,
	height=8.6cm,
	at={(1.453in,1.076in)},
	separate axis lines,
	every outer x axis line/.append style={black},
	every x tick label/.append style={font=\small},
	every x tick/.append style={black},
	xmin=0,
	xmax=40,
	xtick={0,8,16,24,32,40},
	xlabel={$E_{b}/N_{0} \mathrm{(dB)}$},
	every outer y axis line/.append style={black},
	every y tick label/.append style={font=\small},
	every y tick/.append style={black},
	ymode=log,
	ymin=1e-009,
	ymax=1,
	yminorticks=true,
	ytick={1,1e-2,1e-4,1e-6,1e-8},
	ylabel={$\mathrm{BER}$},
	axis background/.style={fill=white},
	xmajorgrids,
	ymajorgrids,
	yminorgrids,
	ylabel near ticks,
	legend style={at={(0.5771,0.745)}, anchor=south west, legend cell align=left, align=left, draw=black}
	]
	\addplot [color=blue, line width=0.7pt]
	table[row sep=crcr]{%
		0	0.115218079591917\\
		0.5	0.108102924207643\\
		1	0.100869862865697\\
		1.5	0.0935725208897955\\
		2	0.0862700971022766\\
		2.5	0.0790258141054329\\
		3	0.0719049211509179\\
		3.5	0.0649723390973165\\
		4	0.0582900960657043\\
		4.5	0.0519147535247344\\
		5	0.0458950539028901\\
		5.5	0.0402700216075079\\
		6	0.0350677128910483\\
		6.5	0.0303047369973778\\
		7	0.0259865711775057\\
		7.5	0.022108583680742\\
		8	0.0186575847544629\\
		8.5	0.0156136677477084\\
		9	0.0129520939382622\\
		9.5	0.010645015390922\\
		10	0.0086629054232913\\
		10.5	0.00697565222339843\\
		11	0.00555334347991103\\
		11.5	0.00436681232535773\\
		12	0.00338802384372976\\
		12.5	0.00259036490716281\\
		13	0.00194887282332004\\
		13.5	0.00144041460460933\\
		14	0.00104381674866078\\
		14.5	0.000739944628214789\\
		15	0.000511734401489909\\
		15.5	0.000344182654251686\\
		16	0.000224298569132964\\
		16.5	0.000141022914160637\\
		17	8.51193813140854e-005\\
		17.5	4.90452096896976e-005\\
		18	2.68066968134629e-005\\
		18.5	1.38008415130202e-005\\
		19	6.64042714861392e-006\\
		19.5	2.96061311306026e-006\\
		20	1.21157537914005e-006\\
		20.5	4.50364214813697e-007\\
		21	1.50312470240528e-007\\
		21.5	4.44683881681184e-008\\
		22	1.14943120291423e-008\\
		22.5	2.55432214870235e-009\\
		23	4.79216084866804e-010\\
		23.5	7.43595589212619e-011\\
		24	9.32442671287342e-012\\
		24.5	9.20472312440711e-013\\
		25	6.94477831399473e-014\\
		25.5	3.8732913245299e-015\\
		26	1.53801867899286e-016\\
		26.5	4.16823122867492e-018\\
		27	7.35250694397989e-020\\
		27.5	8.00351010016099e-022\\
		28	5.06472586023109e-024\\
		28.5	1.74265772719201e-026\\
		29	3.02492662725027e-029\\
		29.5	2.43570992526394e-032\\
		30	8.2817130168066e-036\\
		30.5	1.07014490200611e-039\\
		31	4.66969595702892e-044\\
		31.5	6.0271999645341e-049\\
		32	1.98322115048002e-054\\
		32.5	1.40809278773684e-060\\
		33	1.7891083126414e-067\\
		33.5	3.29771364820893e-075\\
		34	6.96723855081382e-084\\
		34.5	1.29538092452248e-093\\
		35	1.57556046340238e-104\\
		35.5	8.98814717859572e-117\\
		36	1.65564481618325e-130\\
		36.5	6.47748786430409e-146\\
		37	3.36411799236594e-163\\
		37.5	1.3687919428622e-182\\
		38	2.41455101381503e-204\\
		38.5	9.50701184856799e-229\\
		39	3.96693739612896e-256\\
		39.5	7.60486553916562e-287\\
		40	0\\
	};
	\addlegendentry{$\mathrm{Theo.\; OFDM}$}
	
	\addplot [color=blue, line width=0.7pt, forget plot]
	table[row sep=crcr]{%
		0	0.137277462153963\\
		0.5	0.131578849564975\\
		1	0.125755073970784\\
		1.5	0.119834574170088\\
		2	0.113848398252674\\
		2.5	0.107829586565915\\
		3	0.101812448053183\\
		3.5	0.0958317617898723\\
		4	0.0899219423319502\\
		4.5	0.0841162112060208\\
		5	0.0784458169788627\\
		5.5	0.0729393427739489\\
		6	0.0676221332557534\\
		6.5	0.0625158637975557\\
		7	0.057638263880838\\
		7.5	0.0530029959386105\\
		8	0.0486196809600698\\
		8.5	0.0444940540920838\\
		9	0.0406282277464947\\
		9.5	0.0370210365384555\\
		10	0.0336684376081558\\
		10.5	0.0305639411441009\\
		11	0.0276990487165023\\
		11.5	0.0250636807848377\\
		12	0.0226465789384336\\
		12.5	0.020435672624038\\
		13	0.0184184039842865\\
		13.5	0.0165820077676352\\
		14	0.0149137459709429\\
		14.5	0.0134010989217375\\
		15	0.012031915939508\\
		15.5	0.0107945296122082\\
		16	0.00967783818131021\\
		16.5	0.00867136064520957\\
		17	0.00776526905912099\\
		17.5	0.00695040221036849\\
		18	0.00621826444725459\\
		18.5	0.00556101298896041\\
		19	0.00497143658116054\\
		19.5	0.00444292791366226\\
		20	0.00396945179941514\\
		20.5	0.00354551073841737\\
		21	0.00316610915971291\\
		21.5	0.00282671735037871\\
		22	0.00252323584018484\\
		22.5	0.0022519608109865\\
		23	0.00200955093658502\\
		23.5	0.00179299592719754\\
		24	0.00159958694827211\\
		24.5	0.00142688900191693\\
		25	0.00127271529680696\\
		25.5	0.00113510358566014\\
		26	0.00101229441526755\\
		26.5	0.000902711210083225\\
		27	0.000804942094402929\\
		27.5	0.00071772334841913\\
		28	0.000639924388486508\\
		28.5	0.00057053416059091\\
		29	0.000508648837342054\\
		29.5	0.000453460712055852\\
		30	0.000404248188064431\\
		30.5	0.000360366766831969\\
		31	0.000321240944408304\\
		31.5	0.000286356931952234\\
		32	0.000255256122303308\\
		32.5	0.000227529230728494\\
		33	0.000202811043916514\\
		33.5	0.000180775716966305\\
		34	0.00016113256347434\\
		34.5	0.000143622288840697\\
		35	0.000128013621577096\\
		35.5	0.000114100301710014\\
		36	0.000101698389336376\\
		36.5	9.06438600207716e-005\\
		37	8.07904570385679e-005\\
		37.5	7.20077734861097e-005\\
		38	6.41795400184709e-005\\
		38.5	5.72020964558997e-005\\
		39	5.09830277430429e-005\\
		39.5	4.54399467690069e-005\\
		40	4.04994083804123e-005\\
	};
	
	\addplot [color=blue, line width=0.7pt, forget plot]
	table[row sep=crcr]{%
		0	0.119794504193927\\
		0.5	0.112198148698813\\
		1	0.10434010127687\\
		1.5	0.0962680585905056\\
		2	0.088041525140546\\
		2.5	0.0797319644102755\\
		3	0.0714222848030345\\
		3.5	0.0632054878369083\\
		4	0.0551823338690384\\
		4.5	0.0474579416867391\\
		5	0.0401373381154325\\
		5.5	0.0333201122244106\\
		6	0.0270944966036963\\
		6.5	0.0215313745881568\\
		7	0.0166788638375193\\
		7.5	0.012558210715138\\
		8	0.0091617027756101\\
		8.5	0.00645313664245699\\
		9	0.0043710610363824\\
		9.5	0.00283458463734947\\
		10	0.00175107357350247\\
		10.5	0.00102467311570295\\
		11	0.000564387213495046\\
		11.5	0.000290520124505667\\
		12	0.000138639462580636\\
		12.5	6.0781859182625e-005\\
		13	2.42331981899578e-005\\
		13.5	8.68601540258801e-006\\
		14	2.76320036636932e-006\\
		14.5	7.68965944430155e-007\\
		15	1.84185517185131e-007\\
		15.5	3.72859114421329e-008\\
		16	6.25020078867695e-009\\
		16.5	8.47883069546848e-010\\
		17	9.07162538874216e-011\\
		17.5	7.43684436017829e-012\\
		18	4.5223090048164e-013\\
	};
	
	\addplot [color=black , only marks, line width=0.7pt, draw=none, mark size=2pt, mark=square, mark options={solid, black}]
	table[row sep=crcr]{%
		-10	0.205403645833333\\
		-8	0.196940104166667\\
		-6	0.1796875\\
		-4	0.163411458333333\\
		-2	0.141845703125\\
		0	0.1181640625\\
		2	0.0856119791666667\\
		4	0.0567491319444444\\
		6	0.035400390625\\
		8	0.0188176081730769\\
		10	0.00851966594827586\\
		12	0.00338766163793103\\
		14	0.000975587886726547\\
		16	0.000213184232026144\\
		18	2.4567455704951e-005\\
		20	1.07982911489585e-006\\
		22	9.47499951488003e-009\\
		24	0\\
		26	0\\
		28	0\\
		30	0\\
		32	0\\
		34	0\\
		36	0\\
		38	0\\
		40	0\\
	};
	\addlegendentry{$\mathrm{Linear\; GFDM}$}
	
	\addplot [color=black , only marks, line width=0.7pt, draw=none, mark size=2pt, mark=square, mark options={solid, black}, forget plot]
	table[row sep=crcr]{%
		-10	0.211934840425532\\
		-8	0.197041015625\\
		-6	0.186210200471698\\
		-4	0.175641741071429\\
		-2	0.166619769597458\\
		0	0.144538430606618\\
		2	0.117916980421687\\
		4	0.0861730742872807\\
		6	0.0680168999565972\\
		8	0.047264209692029\\
		10	0.0344375825264085\\
		12	0.0192673662475345\\
		14	0.0140925155573593\\
		16	0.00999319275582574\\
		18	0.00596419705124318\\
		20	0.00354482265965167\\
		22	0.00304821407890673\\
		24	0.00160508621220835\\
		26	0.00115294830504749\\
		28	0.000688859662805821\\
		30	0.00043699578752123\\
		32	0.000267568194034998\\
		34	0.000152120462824022\\
		36	0.000107722794705364\\
		38	6.53414001232449e-005\\
		40	4.19058008679747e-005\\
	};
	
	\addplot [color=black , only marks, line width=0.7pt, draw=none, mark size=2pt, mark=square, mark options={solid, black}, forget plot]
	table[row sep=crcr]{%
		-10	0.208577473958333\\
		-8	0.198279747596154\\
		-6	0.184395926339286\\
		-4	0.169173177083333\\
		-2	0.146455652573529\\
		0	0.119280133928571\\
		2	0.0873181573275862\\
		4	0.0550021701388889\\
		6	0.0271809895833333\\
		8	0.00913523204291045\\
		10	0.00177698863636364\\
		12	0.000141879875152519\\
		14	2.91749253121912e-006\\
		16	7.49998080004915e-009\\
		18	0\\
		20	0\\
		22	0\\
		24	0\\
		26	0\\
		28	0\\
		30	0\\
		32	0\\
		34	0\\
		36	0\\
		38	0\\
		40	0\\
	};
	
	\addplot [color=red, only marks, line width=0.7pt, draw=none, mark size=2.0pt, mark=o, mark options={solid, red}]
	table[row sep=crcr]{%
		0	0.13629150390625\\
		2	0.115751378676471\\
		4	0.0844221443965517\\
		6	0.0707691659172662\\
		8	0.0468055761255924\\
		10	0.0286797218658358\\
		12	0.0220304216657366\\
		14	0.015349911971831\\
		16	0.00932156369161106\\
		18	0.00610682294873768\\
		20	0.00373398124284078\\
		22	0.00307356156643082\\
		24	0.0015625624900016\\
		26	0.00113081010616755\\
		28	0.000692996925332633\\
		30	0.00043380750772286\\
		32	0.000270468654875889\\
		34	0.000153304929617161\\
		36	0.000107862757359021\\
		38	6.53250916821419e-005\\
		40	4.30514059488078e-005\\
	};
	\addlegendentry{$\mathrm{FBMC \, OQAM}$}
	
	\addplot [color=red, only marks, line width=0.7pt, draw=none, mark size=2.0pt, mark=o, mark options={solid, red}, forget plot]
	table[row sep=crcr]{%
		0	0.1173828125\\
		2	0.0851236979166667\\
		4	0.0598958333333333\\
		6	0.0335611979166667\\
		8	0.0184280960648148\\
		10	0.00897549715909091\\
		12	0.00330909522804054\\
		14	0.00105338261045259\\
		16	0.000214372601425439\\
		18	2.54912658417649e-005\\
		20	1.13831461933485e-006\\
		22	1.14110544511e-008\\
		24	0\\
	};
	
	\addplot [color=red, only marks, line width=0.7pt, draw=none, mark size=2.0pt, mark=o, mark options={solid, red}, forget plot]
	table[row sep=crcr]{%
		0	0.120703125\\
		2	0.0944010416666667\\
		4	0.0543619791666667\\
		6	0.0260587993421053\\
		8	0.00808465676229508\\
		10	0.00182923274253731\\
		12	0.000128346750393494\\
		14	2.69212000033015e-006\\
		16	6.25020078867695e-009\\
	};
	
	\addplot [color=teal, only marks, line width=0.7pt, draw=none, mark size=2.0pt, mark=x, mark options={solid, teal}]
	table[row sep=crcr]{%
		0	0.134049925085616\\
		2	0.11467879398827\\
		4	0.0913952978971963\\
		6	0.0708644701086956\\
		8	0.0493953203914141\\
		10	0.035051149103139\\
		12	0.0221597753684807\\
		14	0.0145748858128262\\
		16	0.0101365869875454\\
		18	0.00600982328182657\\
		20	0.00375528834613538\\
		22	0.00278572985621752\\
		24	0.0016482324803814\\
		26	0.00109084398509334\\
		28	0.00064622239555856\\
		30	0.000391847815215978\\
		32	0.000270595470568293\\
		34	0.000171488758033504\\
		36	0.000109616310372296\\
		38	6.90664925494508e-005\\
		40	3.74646809724206e-005\\
	};
	\addlegendentry{$\mathrm{OFDM}$}
	
	\addplot [color=teal, only marks, line width=0.7pt, draw=none, mark size=2.0pt, mark=x, mark options={solid, teal}]
	table[row sep=crcr]{%
		0	0.119794504193927\\
		2	0.088041525140546\\
		4	0.0551823338690384\\
		6	0.0270944966036963\\
		8	0.0091617227756101\\
		10	0.00175107357350247\\
		12	0.000138639462580636\\
		14	2.76320036636932e-006\\
		16	6.25021078867695e-009\\
		18	0\\
		20	0\\
		22	0\\
		24	0\\
		26	0\\
		28	0\\
		30	0\\
		32	0\\
		34	0\\
		36	0\\
		38	0\\
		40	0\\
	};
	
	\addplot [color=teal, only marks, line width=0.7pt, draw=none, mark size=2.0pt, mark=x, mark options={solid, teal}]
	table[row sep=crcr]{%
		0	0.115113740808824\\
		2	0.086598624170354\\
		4	0.0586609983158683\\
		6	0.0352567266471119\\
		8	0.0190021203185798\\
		10	0.008893916344404\\
		12	0.00349286065427243\\
		14	0.00107996387053522\\
		16	0.000235118841630757\\
		18	2.89170968390243e-005\\
		20	1.3397921303902e-006\\
		22	1.14665465204211e-008\\
		24	0\\
	};

	\end{axis}
	\spy[black] on (5.495,8.05) in node at (4.4,5.2);
	\node at (5.8,4.3) {$\mathrm{AWGN}$};
	\draw [->] (5.8,4.5)--(6.15,4.9);
	\node at (7.01,4.3) {$\mathrm{TIFS}$};
	\draw [->] (7,4.5)--(7.15,4.85);
	\node at (9.7,6) {$\mathrm{TVFS}$};	
	\draw [->] (9.7,6.2)--(9.8,6.5);
	
	\end{tikzpicture}%
	\caption{BER performance from Linear GFDM, FBMC-OQAM and OFDM under different channel models.}
	\label{bers}
	\end{figure}

	Figure \ref{bers} shows the simulation results regarding the BER performance under the channel models. 
	It is possible to observe that linear GFDM holds the same performance as OFDM and FBMC-OQAM. 
	This performance is observed since no ISI occurs due to the half Nyquist pulse shaping filter and no ICI occurs due to the OQAM configuration.
	For this reason, the system is considered (quasi-)orthogonal, and the BER follows the OFDM bit error probability when the CP is not taken into account.

	For the TIFS channel, the signal-to-noise ratio (SNR) is pondered by the channel's frequency response over the $k$th subcarrier in all waveforms. 
	Thus, a performance degradation is observed when compared with the AWGN channel.
	At the receiver, perfect channel estimation is assumed.
	To perform the channel equalization, the frequency-domain zero-forcing solution was employed \cite{ofdm_matlab}.

	For the TVFS channel, the impulse response changes at every multicarrier frame transmission, thus fashioning a time-variant block fading channel.
	Each subcarrier experiences a flat frequency response, since the channel's coherence bandwidth is approximately $20$ times larger than the subcarrier bandwidth.
	Channel state information is known by the receiver, and the equalization is performed in frequency-domain using  the zero-forcing solution.  	
		
\subsection{Spectral containment performance}

\begin{figure}[h]
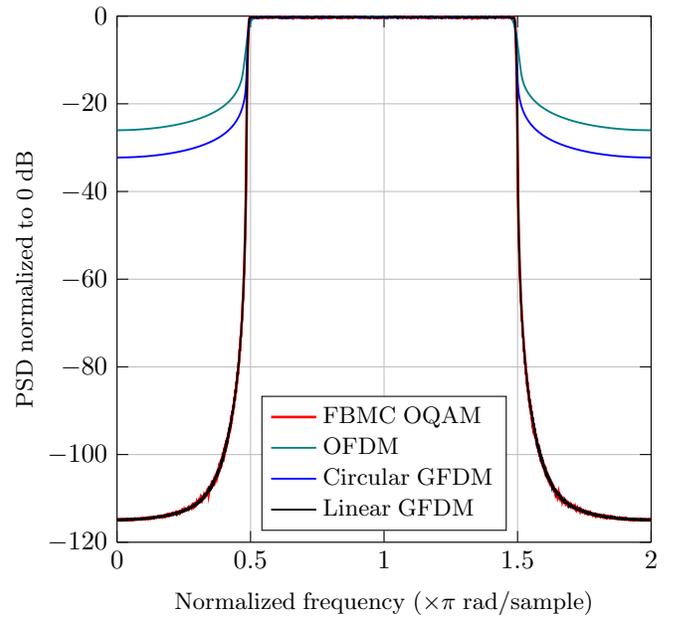

		\centering

		\caption{Power spectral density comparison between Linear GFDM, FBMC-OQAM, circular time-domain GFDM-OQAM and OFDM.}
		\label{psd}
	\end{figure}

	For evaluating the spectrum containment, the power spectral density (PSD) was estimated through the Welch method described in \cite{welch}. We used a set of $10^{3}$ multicarrier frames for the PSD estimation.
	Figure \ref{psd} shows that Linear GFDM holds the same spectrum containment as FBMC-OQAM due to the linear filtering behavior shown by the modified modulation matrices. 
	For the sake of comparison, the OFDM and circular GFDM-OQAM PSDs are also shown. Circular GFDM shows poor spectral containment due to abrupt transitions between the multicarrier frames.
	The PSD is estimated through the baseband signal vector, not taking into account RF front-end impairments.
	
\subsection{PAPR}
	
\begin{figure}[h]
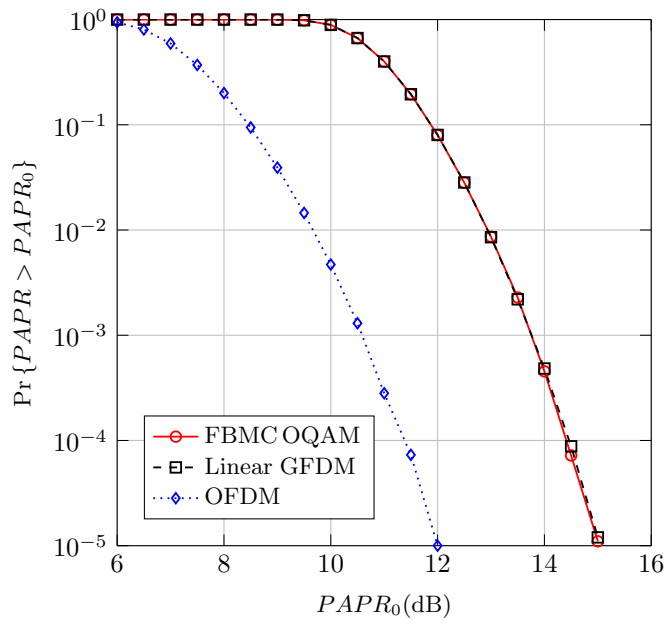

			\centering
			%
%
			\caption{Estimated CCDF of the PAPR for Linear GFDM, FBMC-OQAM and OFDM. }
			\label{papr}
		\end{figure}

	Figure \ref{papr} shows the CCDF of the PAPR from the considered waveforms.

	For estimating the CCDF, we used a set of $10^{6}$ frames.
	Linear GFDM and FBMC show the same poor performance when compared to OFDM due the long filter lengths that are used to shape the data symbols in both schemes.
	
\section{Conclusion}

	In this paper a modification in the GFDM transmit matrix was explored in order to achieve a linear filtering behavior. 
	In fact, linear filtering can provide very good OOB emission performance when compared to circular filtering. 
	Spectral efficiency it is not compromised for achieving the OOB emission performance, assuming a continuous stream of data. 
	Moreover, differently from \cite{gfdmfor5g} no additional techniques, such as windowing, are needed in order to improve spectral containment. 
	Nevertheless, it is important to point out that a longer prototype filter may lead to an increase in latency since it is directly proportional to the transmit signal length.
	Also, longer transmit signals may lead poor performance in fast time-varying channels due to estimation errors. 
	
	As an concept illustration, the future network control can switch from the linear to the circular configuration depending on the user requirements, e.g., deploy Linear GFDM in the RAA use cases and circular GFDM in URLL use cases.
	As shown, GFDM can be configured to perform like FBMC bearing all its characteristics. 
	Hence, the GFDM's flexibility shows its feasibility to be deployed as a configurable software defined waveform for the 5G radio access network.
	
\bibliographystyle{IEEEtran}
\bibliography{IEEEabrv,my_references}
	
\end{document}